\documentclass{article}

\usepackage{arxiv_preprint}
\usepackage{times}

\usepackage[utf8]{inputenc}
\usepackage[T1]{fontenc}
\usepackage{hyperref}
\usepackage{url}
\usepackage{booktabs}
\usepackage{amsfonts}
\usepackage{nicefrac}
\usepackage{microtype}
\usepackage{xcolor}
\usepackage{graphicx}
\usepackage{multirow}
\usepackage{array}
\usepackage{makecell}
\usepackage{enumitem}
\usepackage{caption}
\usepackage{subcaption}
\usepackage{xspace}
\usepackage{float}
\usepackage{tabularx}
\usepackage{amsmath, amssymb}
\hypersetup{
  colorlinks=true,
  linkcolor=blue!60!black,
  citecolor=blue!60!black,
  urlcolor=blue!60!black,
}

\newcommand{\bench}{AgenticVBench\xspace}

\title{\bench: Can AI Agents Complete Real-World Post-Production Tasks?}

\author{%
  Zongheng Cao, Yi Zheng, Rui Song, Xinyu Hu \\
  Philo Labs Research \\
  \texttt{research@philolabs.ai}
}

\begin{document}

\maketitle

\begin{abstract}
Video production workflows offer a rich and demanding arena for evaluating multimodal AI agents: they require composite capabilities across text, image, audio, and video understanding, along with long-horizon planning, and tool use. To this end, we introduce \textbf{AgenticVBench}, a benchmark of 100 agentic tasks across 4 task families spanning the real world post-production workflow, constructed from real production workflows contributed by 20 industry experts averaging 6 years of professional experience. Both the tasks and their evaluation specifications are authored by these experts, combining programmatic verifiers and expert rubrics. We evaluate frontier vision-language models (VLMs) with both vendor-native and open-source harnesses. The best evaluated agent stack barely crosses 30\%, far below human expert performance on the same tasks. We further find that the choice of harness substantially affects model behavior, including scores, tool-use patterns, and failure modes. AgenticVBench provides a foundation for diagnosing and improving both models and harnesses for agentic video production. Benchmark website: \url{https://agenticvbench.com}.
\end{abstract}

\section{Introduction}
\label{sec:intro}



Video production, a domain of substantial economic and cultural value, is increasingly empowered by AI agents, from creator tools and general-purpose video agents~\citep{liang2025univa, li2026direct} to professional studio post-production. State-of-the-art AI agents are moving beyond single-shot question answering (QA) to long-horizon tasks that combine visual and audio perception~\citep{sakshi2024mmau}, semantic and narrative reasoning~\citep{fu2026video}, retrieval over long video contexts~\citep{zhou2025mlvu}, and editing and production workflows. However, current multimodal benchmarks focus on perception and reasoning through single-shot QA settings~\citep{fu2025videomme, yue2024mmmu,huang2024vbench, yang2024swe}, while existing agentic benchmarks focus mainly on software engineering rather than multimodal video-production workflows~\citep{zhou2023webarena,koh2024visualwebarena,xie2024osworld,jimenez2024swebench,merrill2026terminalbench}. None effectively measure the composite capabilities demanded by real-world video production workflows. This gap hinders systematic model and harness design for video agents.
\begin{figure}[h]
    \centering
    \includegraphics[width=0.99\linewidth]{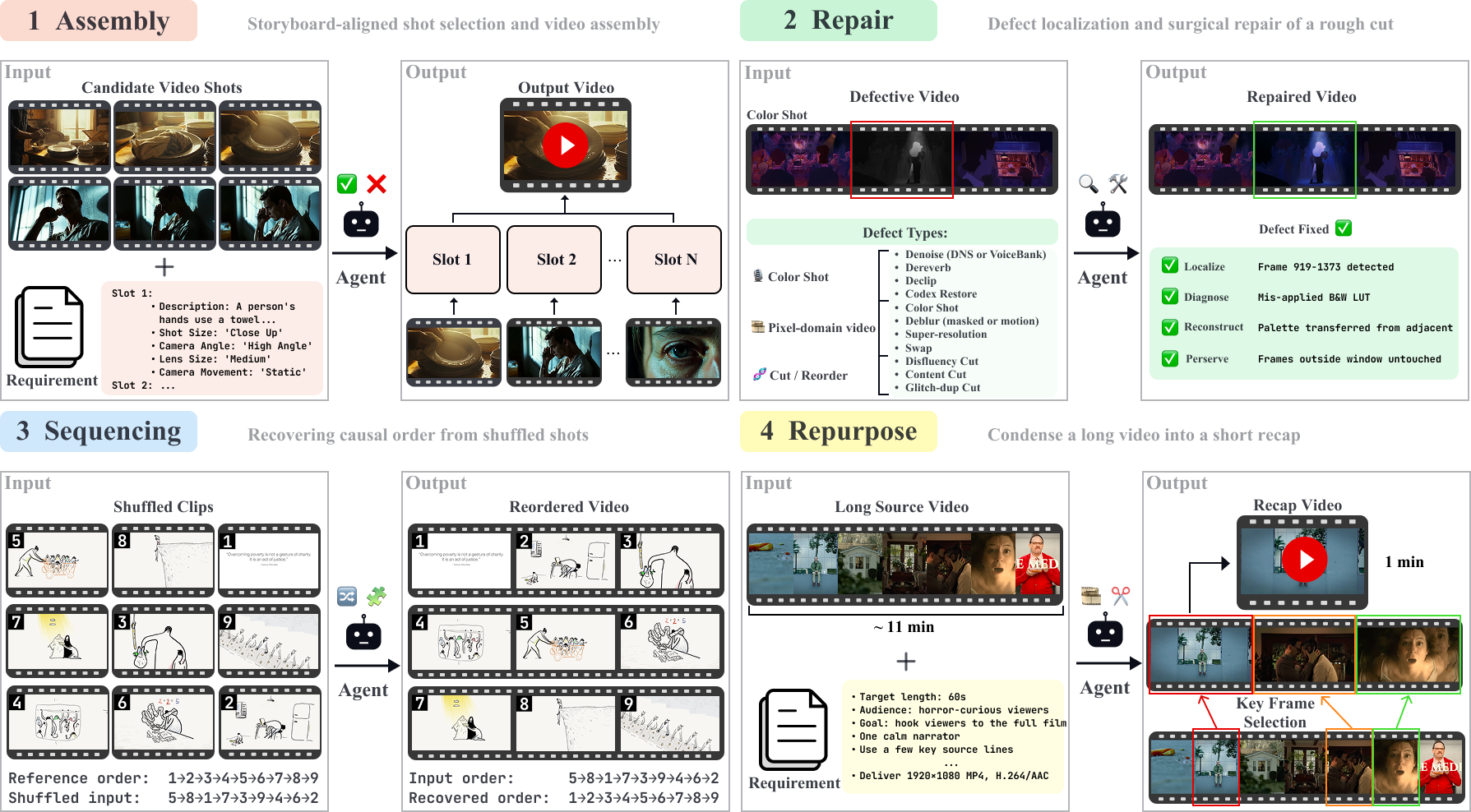}
    \caption{Overview of the four task families in AgenticVBench. Assembly evaluates storyboard-aligned shot selection and video construction, Repair focuses on defect localization and correction, Sequencing tests causal and temporal shot ordering, and Repurpose measures long video condensation into short-form deliverables such as recap, trailer, teaser, and social cut.} 
    \label{fig:overall_fig}
\end{figure}

To address this gap, we introduce AgenticVBench, a benchmark evaluating video agents on four task families that span post-production: Assembly, selecting clips that match a storyboard; Repair, finding and fixing technical defects; Sequencing, restoring the order of shuffled shots; and Repurpose, turning long source videos into brief-driven short deliverables. Figure \ref{fig:overall_fig} shows the details of the four task families in AgenticVBench. The task families are authored by 20 industry experts from various scenarios of video production, including traditional film studios, AI film studios, video agent companies, and creators. Each task is equipped with verifiers co-authored by industry experts, including both objective and subjective items.


AgenticVBench offers multifaceted advantages over existing multimodal benchmarks. First, \textbf{task family coverage}: it is the first benchmark to systematically cover four major task families of video agents across the full multimodal stack of text, image, video, and audio. Second, \textbf{production realism}: tasks are authored by industry experts based on professional production experience, with rigorous quality control. Third, \textbf{open evaluation specification}: we open-source full rubrics alongside the benchmark. Lastly, \textbf{long-horizon difficulty}: tasks span 0.5 hours to one week of human completion time.

We evaluate 7 state-of-the-art frontier vision-language models (VLMs) with their native harnesses (Claude Code CLI, Codex CLI, Gemini CLI) and open-source harnesses (OpenClaw, OpenCode). The best model-harness combination scores only 31\%, substantially lower than human performance on the same tasks. We further find that the choice of harness matters, with the maximum performance difference reaching 20pp.

\section{Benchmark Construction}
\label{sec:bench}

\subsection{Expert Pool}
We recruited 20 industry experts from four distinct production contexts. The four contexts and their respective expert backgrounds are described in Table~\ref{tab:expert_pool} in Appendix~\ref{app:recruitment}. All experts went through a standardized selection process: portfolio review, video interview, and calibration training module for task validation. The experts involved received industry-standard compensation. Collectively, these 20 experts defined the four task families, and their workflows determined the division of labor during task and rubric authoring.

\subsection{Task Family Framework}

The four task families are designed through a multi-step bottom-up pipeline. Each expert was asked to independently draft a set of end-to-end task briefs based on their daily workflows, along with their deliverable time windows and requirements. Our team reviewed the submission with the experts and identified the task families that (i) correspond to long-horizon production tasks with video production realism, and (ii) admit verifiable evaluation, either through programmatic tests or through atomic yes/no rubrics. The second criterion rules out, for example, aesthetic evaluation without reference, ideation without structured requirements, or other dimensions that lack reproducible measurement. Under these two criteria, four task families that recurred across all expert contexts were retained. The full task authoring framework is described in Section~\ref{app:recruitment}.

This bottom-up framework aligns with established conventions in video editing scholarship and anchors each task family in a common stage of post-production~\citep{murch2001blink, dmytryk1984film, pearlman2009cutting}. \textbf{Assembly} maps to rough-cut construction, drawing on shot selection and continuity as editors select takes against a storyboard and form the first edit. \textbf{Repair} maps to review and finishing, corresponding to the physical rhythm layer as editors localize defects in an existing cut and apply targeted corrections. \textbf{Sequencing} maps to narrative reorganization, requiring story and event-rhythm reasoning to restore temporal and causal order across shuffled shots. \textbf{Repurpose} maps to end-to-end compression, adding emotion and holistic editorial judgment as editors turn long source material and a client brief into a short deliverable.

These four task families are not designed to be orthogonal. They instead provide workflow-oriented coverage of post-production, from rough construction to final deliverable. The experimental results in Section~\ref{sec:experiments} show that frontier models and harnesses exhibit distinct failure modes across each task family.

\subsection{Task Creation}

Once the task families are established, task creation diverges into two paths depending on the task family.

\textbf{Programmatic task families} (Assembly, Repair, Sequencing). These task families have systematic structure: Assembly consists of a storyboard plus candidate selection; Repair consists of a source video with an injected defect; Sequencing consists of shuffled shots with a story overview. Source video selection criteria are jointly determined by experts and the project team. Distractor generation and defect injection protocols are designed by the team and reviewed by experts. Once protocols are finalized, we generate tasks in batches using publicly accessible source videos; each instance undergoes expert review as part of quality control.

\textbf{Expert-authored task family} (Repurpose). Repurpose tasks are open-ended, the brief is like a realistic customer request and the deliverable has no single correct answer, making automatic task creation infeasible. Repurpose tasks are authored by experts, including the task brief, source video selection, and the corresponding rubrics covering format, visual, narrative and sound pillars.

Rubrics follow a similar dual-track design, each track corresponds to a different verification method:

\textbf{Programmatic verifiers} are used for Assembly, Repair, Sequencing, and the Format pillar of Repurpose. Experts provide domain-specific standards, and the team operationalizes these into deterministic checks.

\textbf{Expert rubrics} are used for the Visual, Narrative, and Sound pillars of Repurpose. These are authored by experts; the team performs formatting and atomicity review. All rubric items are binary (Yes/No), so grading does not rely on the grader's free-form judgment.

All source videos come from public channels. In the entire task creation process, experts contribute workflow knowledge - task types, deliverable standards, and project timeline from daily production - rather than specific client projects from their own portfolios. Each task instance is built independently from publicly accessible source videos and contains no proprietary client material.

\subsection{Task Family: Assembly}

\textbf{Task Definition.}
For each task in this family, there is a storyboard with $3$-$6$ slots. Each slot is defined according to a shot description and four cinematic variables: shot size, camera angle, lens size, and camera movement~\citep{liu2025shotbench}. The agent receives a shuffled set of candidate clips. For each slot, there is one golden clip extracted from the original source video and a few AI-generated distractors. The agent is asked to select the candidate clip that best matches each storyboard slot, then assemble the selected clips, and finally submit the integrated video. 

\textbf{Distractor Design.}
We create each distractor by modifying the original source slot along exactly one cinematic dimension. This design requires the agent to inspect fine-grained cinematic dimensions, rather than relying on coarse-grained video content matching. Distractors are generated by first regenerating the initial frame using Nano Banana Pro~\citep{google2025nanobananapro}, followed by image-to-video generation using Seedance 2.0~\citep{seedance2026seedance}. We discard audio from candidate clips because current AI video regeneration cannot preserve the original audio, and hence could be exploited by agents to get the result without analysis. 

\textbf{Source Video Selection.}
We take submissions from the 2025 Runway AI Film Festival as source material for two considerations. First, the shortlisted films cover a wide range of visual styles, including realistic, cartoon, anime, narrative, and abstract forms. In addition, because the source videos are AI-generated, the distractors mimic the golden clips in a more natural way, reducing cues that could make the task easier.

\textbf{Evaluation.}
Each task is scored with adjustment for randomness. $r$ is the proportion of storyboard slots for which the agent selects the correct clip, and let $k$ be the number of candidate clips per slot. The score is
\begin{equation}
  \mathrm{score} = \frac{r - 1/k}{1 - 1/k}
\end{equation}
A score of $0$ corresponds to random selection, and a score of $1$ corresponds to selecting the correct clip for every slot.

\textbf{Coverage.} $18$ stories, from $9$ source films, with an average of $4.00$ storyboard slots per story. Each slot has 1 golden and 2 distractor variants, and a total of $216$ candidate clips in the task family.

\subsection{Task Family: Repair}
\textbf{Task Definition.} Inputs include a broken video with an injected defect and a repair prompt. The prompt includes the client request style issues in the broken video but does not explicitly name the precise timestamp or defect type. The agent needs to localize and identify the defect, submit a fixed video and generate a report; the report should include the reasoning process including the type and timestamps.

\textbf{Defect Families.} Corruptions span three groups: audio defects
(background noise, room echoes, loud-audio distortion), visual defects
(color shifts, blurriness, lost sharpness), and timeline defects
(out-of-order shots, spoken filler words, repeated frames, and
factually-wrong sentences). Each corruption affects one time window
inside the clip; the rest is clean.

\textbf{Source Video Selection.} All source clips are human-authored
professional video, drawn from movies and animated films, broadcast
news, TV sketch, music video, ads, and YouTube creator content (tech
reviews, vlogs, educational, interviews).

\textbf{Evaluation.} A programmatic verifier gives each task one
reward between $0$ and $1$. For every measurement we use, the broken
input the agent receives scores $0$ and the bundled golden reference
scores $1$, and the agent's output is placed linearly between the two.
For audio and visual defects, the corruption sits in a known time
window inside the clip; the reward primarily measures how well the
agent fixed that window, with a smaller term that penalizes changes to
the rest of the clip. For timeline defects, the agent reports a list of
time ranges to remove or rearrange, and the reward is the fraction of those
ranges that match the ground truth, gated by a check that the rendered
output actually matches the reported edits. The reward drops to $0$ if
the output is missing, malformed, or a verbatim copy of the broken
input. Full per-family formulas are in Appendix~\ref{app:rubric_repair}.

\textbf{Coverage.} 18 tasks across three groups: audio defects,
visual defects, and timeline defects.

\subsection{Task Family: Sequencing}

\textbf{Task Definition.}
Given a brief story overview and a set of shuffled clips from a source video, the agent is asked to recover the correct narrative order. Two expected outputs: the reordered video, and the predicted clip ordering.

\textbf{Task Construction.}
For each task, we divide a longer video into $7-20$ clips. These clips are randomly shuffled before being provided to the agent, along with a short description ($1$-$2$ sentences of the ordered video. The source pool consists of videos including more than 15 content categories: AI-generated films, commercials, cinematic-shot examples and more. This design tests whether agents can navigate narrative order by integrating multimodal continuity.

\textbf{Evaluation.}
Score is calculated by comparing the predicted clip order to the ground-truth order using three metrics: normalized distance (ND), longest increasing subsequence (LIS), and adjacent fidelity (ADJ). ND measures the normalized total displacement of clips, LIS measures the largest correctly ordered narrative backbone, and ADJ measures whether local clip-to-clip transitions are preserved. The final score is
\begin{equation}
    \mathrm{score}
    =
    (1-\mathrm{ND}) \cdot \mathrm{LIS} \cdot \mathrm{ADJ}.
\end{equation}
Here a multiplicative score is used to penalize random orderings sharply, while rewarding predictions only when they achieve high alignment across three metrics. Detailed formula definitions are provided in Appendix~\ref{app:rubric_seq}.

\textbf{Coverage.} $28$ accepted sequence instances from $27$ source videos, averaging $13.71$ clips per instance.

\subsection{Task Family: Repurpose}

\textbf{Task Definition.} Input consists of a source video with a length of 4 minutes to 3 hours and a creative brief markdown. The brief briefly states the source video's content, audience profile, target platform and the deliverables. It also describes the desired tone, mood, and pacing and specifies ``hard'' delivery requirements: runtime, aspect ratio, container format, and audio inclusion. The brief offers limited creative directions depending on the source video. The agent needs to deliver a standalone short-form repurposed video (e.g., recap, trailer, teaser, or social cut) that follows these constraints while preserving the key content of the source.

\textbf{Source Video Selection.} 
The core challenge of the Repurpose task is \emph{long-horizon compression}: deciding what to retain, discard, and restructure from tens of minutes of source footage. We sample four content types with different editorial demands: talks stress argument compression, narrative shorts stress plot and emotional continuity, sports broadcasts stress action tracking and reframing, and music performances stress rhythmic alignment.

\textbf{Evaluation.} Each task is evaluated with approximately 30 binary rubrics totaling roughly 36 points. The rubric has five pillars: Format checks hard delivery requirements with a programmatic verifier; Visual checks scene preservation, color and lighting continuity, and transition legibility; Narrative checks story arcs, key beats, turns, and reveals; Sound checks audio quality, voice consistency, synchronization, and audio arcs; and optional Penalty items deduct for critical violations that make the deliverable hard to follow or outside the brief.

\textbf{Coverage.}
36 tasks across four content type breakdown, total number of 1,069 rubric items, average ${\sim}30$ items per task.









\subsection{Quality Control}
\label{sec:quality-control}
Each candidate task passes four quality-control(QC) gates before entering the scored benchmark. \textbf{Authoring QC} checks that the brief is a realistic production request, that the expected deliverable is feasible under the task budget, and that each rubric item is observable from the submitted artifacts. \textbf{Asset QC} checks that videos decode cleanly with \texttt{ffprobe}, that prompts and manifests refer only to files present in the task bundle, and that source provenance and license metadata are recorded. \textbf{Expert QC} requires review by the task author and by an independent reviewer from the expert pool or project team; items marked ambiguous, non-atomic, or outside the stated brief are revised or removed. \textbf{Verifier QC} runs golden-path and adversarial submissions through the scorer to check for parser failures, reward leakage, and zero-floor behavior.

The final benchmark contains only accepted instances: 18 Assembly, 18 Repair, 28 Sequencing, and 36 Repurpose tasks. Candidate instances that fail source-readiness, artifact-integrity, or rubric-atomicity checks are not included in any denominator. Appendix~\ref{app:recruitment} describes recruitment and calibration, and Appendix~\ref{app:iaa} describes the agreement statistics used for Repurpose rubric validation.

\section{Experiments and Analysis}

\label{sec:experiments}
We evaluate 7 frontier vision-language models (Claude Opus 4.7, Claude Sonnet 4.6, GPT-5.5, GPT-5.4-mini, Gemini 3.1 Pro, Gemini 3 Flash, and Qwen3-VL-235B-A22B-Instruct) in 20 model-harness combinations on a curated set of 100 tasks (Repurpose 36, Sequencing 28, Repair 18, Assembly 18). Each model runs on two open-source harnesses (OpenCode and OpenClaw); models with first-party scaffolding additionally run on their vendor-native harness (Claude on Claude Code, GPT on Codex, Gemini on Gemini CLI). Qwen has no vendor-native option. Each (model, harness, task) cell runs $K{=}3$ times under a fixed tool schema, identical inputs, matching rollout limits, and shared random seeds (Appendix~\ref{sec:appendix-setup}). Per-task-family scores are normalized to $[0, 1]$. Failed rollouts (cells where the harness produced no usable output) are scored as $0$. We additionally collect expert deliverables on the same tasks as a human-baseline reference (Appendix~\ref{sec:appendix-baseline}).

\subsection{Main Results}
\label{sec:headline}

\begin{figure}[H]
    \centering
    \includegraphics[width=\linewidth]{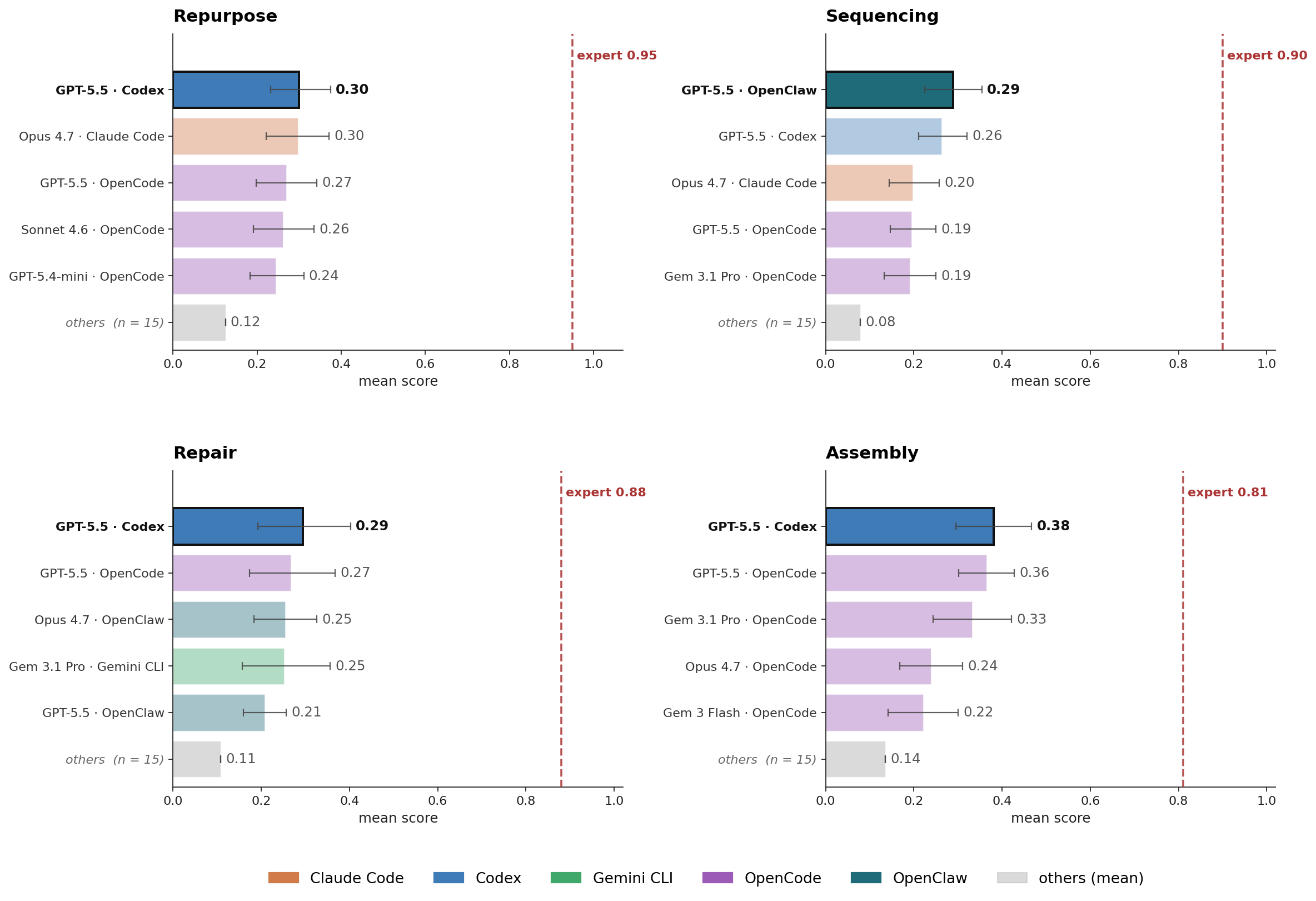}
    \caption{Mean score per (model, harness) combination on four video production task families. Red dashed line: expert human reference. The agent-expert gap is $43$ to $65$ percentage points across families. GPT-5.5 wins every task family; Codex is the winning harness on three of four, with OpenClaw winning Sequencing.}
    \label{fig:headline}
\end{figure}

Agents in our matrix complete the video production tasks. An agent given a 12-minute film and a brief produces a watchable \texttt{.mp4} with cuts, narration, and shot transitions; the best stack on Assembly grounds shots at $0.38$ (roughly half of expert), and the best on Repair localizes defect windows at $0.30$. The headroom is substantial: Repurpose shows the largest gap at $65$ pp (best stack $0.30$ vs expert $0.95$), while Assembly shows the smallest at $43$ pp (best stack $0.38$ vs expert $0.81$). The same model can produce substantially different scores under different harnesses, and the variation across harness, vendor pairing, and modality routing is the subject of next section.

\subsection{Model Behavior}
\label{sec:behavior}
This section looks at two things: how each combo's tool use breaks down, and what failures look like.

\begin{figure}[H]
    \centering
    \includegraphics[width=0.85\linewidth]{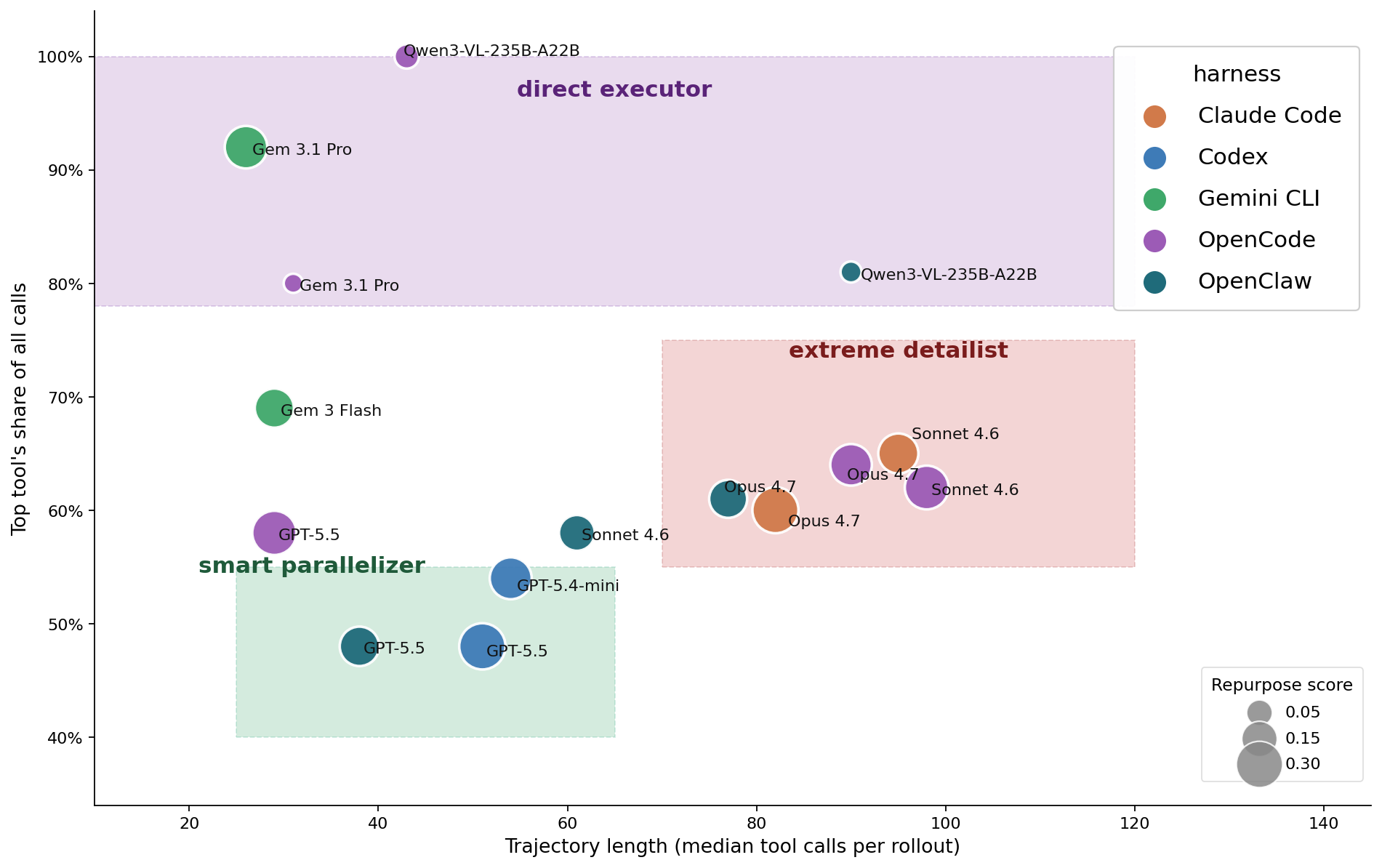}
    \caption{Behavioral signature per (model, harness) combination. X: median tool calls per rollout. Y: share of all calls allocated to the single most-used tool. Bubble size: Repurpose score. Color: harness. Three archetypes separate cleanly. \emph{Smart parallelizer} (low calls, low concentration): the agent spreads work across multiple tools. \emph{Extreme detailist} (high calls, mid-high concentration): the agent over-investigates by repeatedly re-reading the same files. \emph{Direct executor}: the agent locks onto a single tool and never diversifies.}
    \label{fig:behavior}
\end{figure}

\paragraph{Per-model traits.}
Plotting median trajectory length against tool concentration (Figure~\ref{fig:behavior}) and reading individual trajectories together surface three behaviors that recur across (model, harness) cells. \textbf{Smart parallelizers} (GPT-5.5) build composite frame strips and inspect them in a single \texttt{view\_image} call instead of reading frames one-by-one, and bundle several edit operations into a single Python function where other models drive the same operations through multiple shell turns. \textbf{Extreme detailists} (Claude Opus 4.7, Sonnet 4.6) read many frames before writing a plan, then proactively recheck the deliverable and patch their own errors before submission. \textbf{Direct executors} (Gemini 3.1 Pro, Qwen3-VL-235B-A22B-Instruct) start issuing actions before reading enough context, lock onto one tool, and submit without rechecking.

\textbf{Failure mode summary.}
Reading the failed trajectories surfaces four recurring reasons (Table~\ref{tab:failure-modes}). Repurpose fails most often through long-context information loss: agents burn the rollout budget on full-source whisper transcription or repeated frame thumbnailing and never reach the assembly step. Repair's dominant failure is temporal reasoning: agents render a valid \texttt{.mp4} but pick the wrong cut window, with median $\Delta\text{start}$ from $15$ to $100+$ seconds. Modality misalignment (audio defect unfixed; narration over silent films) and hallucinated grounding (wrong scene; fabricated premise) round out the top four. The mix is sharply task-dependent, so reporting failures as a single rate misroutes engineering effort.

\begin{table}[h]
    \centering
    \footnotesize
    \begin{tabular}{p{0.46\linewidth}rr}
        \toprule
        Failure reason & Repurpose (\%) & Repair (\%) \\
        \midrule
        Long-context information loss    & $\mathbf{83}$ &  $0$ \\
        Temporal reasoning                                 &  $1$ & $\mathbf{65}$ \\
        Modality misalignment        & $10$ & $24$ \\
        Hallucinated grounding            &  $6$ & $11$ \\
        \midrule
        $n$ failures                                                         & $153$  & $237$ \\
        \bottomrule
    \end{tabular}
    \vspace{4pt}
    \caption{Top four failure reasons on Repurpose and Repair.}
    \label{tab:failure-modes}
\end{table}

\subsection{Harness as a First-Order Variable}
\label{sec:harness}

End-to-end agent performance is determined jointly by the model and the harness. Two sub-findings make this concrete: holding the model fixed produces a substantial within-model spread, and current harnesses are not multimodal-native at the action layer.

\begin{figure}[H]
    \centering
    \begin{subfigure}[t]{0.48\linewidth}
        \centering
        \includegraphics[width=\linewidth]{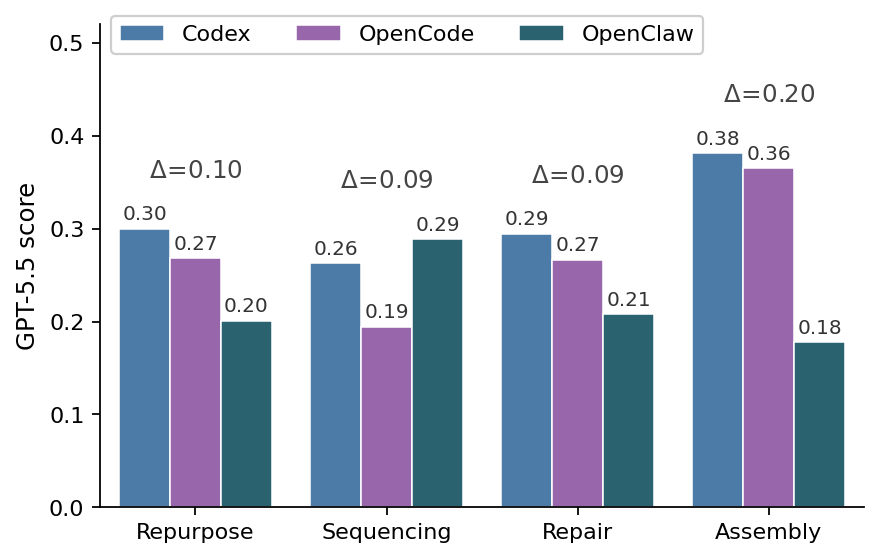}
        \subcaption{GPT-5.5 within-model harness spread.}
        \label{fig:within-model-spread}
    \end{subfigure}
    \hfill
    \begin{subfigure}[t]{0.48\linewidth}
        \centering
        \includegraphics[width=\linewidth]{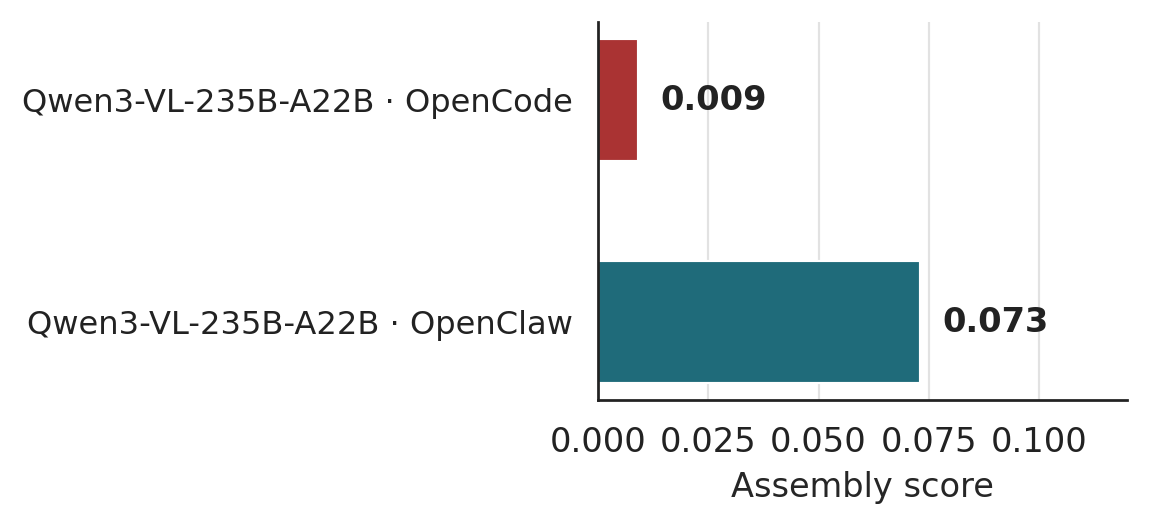}
        \subcaption{Open-source VLM across open-source harnesses.}
        \label{fig:multimodal-routing}
    \end{subfigure}
    \caption{Two harness effects on a fixed model. \textbf{(a)} \textbf{GPT-5.5} across three harnesses (Codex, OpenCode, OpenClaw) on the four task families: within-model harness spread $\Delta$ reaches $0.20$ on Assembly, comparable to the gap between adjacent models on the leaderboard. \textbf{(b)} Same model (Qwen3-VL-235B-A22B-Instruct) on Assembly: $0.009$ on OpenCode versus $0.073$ on OpenClaw. The eight-fold gap on a fixed model points to a harness-level incompatibility between OpenCode and Qwen3-VL that does not appear when OpenCode hosts other models.}
    \label{fig:harness-effects}
\end{figure}

\paragraph{Same model, different harness, substantial spread.}
Holding the model fixed and varying the harness shifts GPT-5.5's Assembly score by $20$ pp ($38$ / $37$ / $18$ on Codex / OpenCode / OpenClaw), comparable to the gap between adjacent models on the leaderboard (Figure~\ref{fig:harness-effects}a). The score pattern reflects how each harness wraps the planner, not just what tools it ships. Codex interleaves reasoning blocks with tool calls (reasoning-to-action ratio $0.4$ on GPT-5.5, $1.0$ on GPT-5.4-mini); paired with GPT-5.5, this deliberative loop wins three of four families (Repurpose, Repair, Assembly). OpenClaw routes modality work to sub-models (vision to a sub-VLM, narration to OpenAI TTS), so the planner can focus on narrative-beat reasoning across many clips; this wins Sequencing, but the planner never sees pixels directly and the harness drops by half on Assembly. Claude Code's Plan and TodoWrite tools force the agent to write and revise a plan as it works, keeping Opus 4.7 competitive across families through deep state-tracking (Appendix~\ref{sec:appendix-trajectory-repair}). OpenCode is cache-first: per-turn input tokens drop close to zero after warmup while cache reads stay at $15$-$27$K, which keeps long \texttt{bash} / \texttt{read} loops stable and fits Repair's \texttt{ffmpeg} pipelines. Gemini CLI is the leanest of the five, with zero planning artifacts and no orchestration of its own, so 3.1 Pro holds $0.12$-$0.25$ across families while Flash drops to $0.03$ on Sequencing with nothing to fall back on.

\paragraph{Harnesses are not multimodal-native.}
The model can only use what the harness exposes. OpenCode paired with Qwen3-VL-235B-A22B scores $0.009$ on Assembly while the same model on OpenClaw scores eight times higher (Figure~\ref{fig:harness-effects}b); the gap is harness-attributable on a fixed model, though we do not have Assembly traces to attribute the precise routing or tool-discovery difference. OpenClaw is the only harness in our matrix that ships typed primitives for \texttt{image}, \texttt{tts}, \texttt{music\_generate}, and \texttt{video\_generate}, routed through harness-managed sub-models. The other four expose only generic shell, file, and (in some cases) view-image tools, leaving the agent to compose multimodal pipelines from scratch. OpenClaw's typed primitives let it win Sequencing through cleaner narrative-beat reasoning over many clips; the sub-model routing also costs pixel detail on Assembly, where OpenClaw drops $20$ pp from the Codex pairing. What looks like a model gap on a given cell is often a harness gap, and typed multimodal primitives are the architectural direction the field has not yet committed to. 


\subsection{Ablation}
\label{sec:ablation}

To probe what limits each task family, for every family we select the top-5 (model, harness) combinations from Section~\ref{sec:experiments} and run each combination $K{=}3$ times. The intervention is either an oracle hint added to the instruction or a single instruction field stripped.

\textbf{Repair: localization and tool use are joint bottlenecks.} Repair decomposes into two sub-skills: \emph{localization} (finding where the defect is) and \emph{tool use} (planning and applying the fix). Providing an oracle defect location isolates the second step and lifts the mean Repair score by $+13$ pp. The remaining gap reflects domain-specific tool use: defects such as color grading, super-resolution, and audio restoration require professional editing operations that the oracle does not address.

\textbf{Sequencing: visual-temporal reasoning is the bottleneck.} Sequencing requires reconstructing narrative order from visual and temporal cues combined with causal reasoning over the source content. Prepending a paragraph-long narrative description simplifies the task to a video-captioning comparison, where the agent only needs to match each clip to its described beat. Even with this simplification, providing the narrative lifts mean Sequencing score by only $+22$ pp, indicating that multimodal visual-temporal reasoning remains a real constraint for current agents.

\textbf{Assembly: agents rely on description, not cinematic detail.} Stripping the per-slot \texttt{description} field drops mean Assembly score by $27$ pp, while stripping \texttt{camera\_movement} moves the mean by $1$ pp. The cinematic-shorthand fields (shot size, camera angle, lens size, camera movement) contribute essentially nothing on top of the prose description. By contrast, human editors treat these cinematic fields as the primary variables when matching candidate clips to a storyboard, revealing a gap between how agents currently solve Assembly and how the task is solved professionally.

\textbf{Repurpose: long video understanding remains challenging.} Repurpose covers a wide range of source video types including films of all genres, music videos, sports broadcasts, and talks, with source durations up to several hours. Long video understanding is the binding challenge for current multimodal agents: prepending an editor's reference document with comprehensive narrative notes for each scene, what happened, and the connections between them yields a $+23$ pp lift on the frontier-floor task \texttt{comedy\_knead}. The lift is most pronounced for films with complex logical cues such as the non-linear narrative in \emph{Mulholland Drive}, where the agent cannot recover the intended structure from the raw source video alone.

\section{Related work}
\label{sec:disc:related}

\textbf{Multimodal benchmarks.} Image and video QA benchmarks test perception and reasoning through single-pass question answering or short-form responses~\citep{yue2024mmmu, fu2025videomme, zhou2025mlvu, zhang2025videoads}. More specialized benchmarks target cinematic understanding~\citep{liu2025shotbench} and audio-visual reasoning~\citep{sakshi2024mmau}. These benchmarks are useful for measuring what a model can understand from images, video, or audio, but they do not require the model to plan over many steps, interact with an editing workspace, call external tools, or produce a finished media artifact. \bench instead evaluates whether multimodal understanding can be converted into concrete production actions.

\textbf{Agentic benchmarks.} Agentic benchmarks measure long-horizon planning and tool use, but existing work mainly targets browser and desktop control~\citep{zhou2023webarena, koh2024visualwebarena, xie2024osworld}, code and command-line workflows~\citep{jimenez2024swebench, merrill2026terminalbench}, and general task assistance~\citep{mialon2023gaia, patwardhan2026gdpval}. They establish the value of evaluating models inside tool-using environments, where success depends on planning, state tracking, and recovery from intermediate errors. However, they do not target video production workflows, where agents must inspect audiovisual material, preserve continuity and narrative structure, manipulate media files, and satisfy both technical and editorial requirements. \bench extends agentic evaluation to this multimodal production setting.


\textbf{Video generation and editing.} Video-generation benchmarks score visual fidelity, temporal consistency, and physical plausibility~\citep{huang2024vbench, wang2024you}. Single-instruction video editing benchmarks evaluate the quality of a particular edit over one $(\text{source video}, \text{prompt})$ pair at a time~\citep{sun2025ve, li2025fivebench, chen2025ivebench, gao2026vefx, chen2025editboard, wu2025veditbench}, situated within a broader video-editing literature~\citep{sun2024videoeditingsurvey, song2023solving}. These benchmarks characterize the quality ceiling of a video model or editing tool, but do not evaluate whether an agent can decide which operations are needed, choose and sequence assets, localize defects, and apply tools across a multi-step production task. In contrast, \bench evaluates end-to-end agentic video production.



\section{Discussion and Conclusion}
\label{sec:disc:future}

\textbf{Limitations and Future work.}
Our evaluation covers frontier models and a focused set of vendor-native and open-source harnesses. Many capable models, editing tools, and agent frameworks are not included because of compute, cost, and access constraints. The benchmark also has scope limits: our tasks emphasize post-production workflows represented by our expert pool and source-video collection, and future versions should broaden coverage across languages, regions, genres, production styles, and professional conventions. In addition, some task families rely on public media assets and evolving model APIs, so maintaining reproducibility will require versioned releases, stable verifier code, and careful documentation of source provenance.

A second limitation is that video production contains both objective and subjective dimensions. Our programmatic verifiers cover mechanically checkable requirements, while expert rubrics capture editorial criteria that are difficult to reduce to deterministic tests. Although this design reflects real production practice, future work should further study rubric sensitivity, inter-rater agreement, source-grounding, and robustness to shallow or format-only solutions. More broadly, a benchmark for video-editing automation has labor-market implications for the same production roles represented in our expert pool. The current gap to expert performance suggests that deployment should remain human-supervised. We release \bench as a diagnostic tool for studying model and harness failures, improving provenance and reproducibility, and supporting safer evaluation of agentic video systems.

\textbf{Conclusion.}
\bench introduces a 100-task expert-authored benchmark for agentic video post-production across Assembly, Repair, Sequencing, and Repurpose. Tasks span 0.5 hours to one week of expert effort and require composite reasoning over text, image, audio, and video, together with long-horizon planning and tool use. Our results show that current frontier agent systems remain far from expert performance: the best stack reaches under half of expert performance, and failures vary substantially across task families. We further find that harness design is a first-order factor, affecting both scores and failure modes. These results suggest that progress in agentic video production will require not only stronger multimodal models, but also better tool interfaces, state management, verification, and workflow-aware harness design.
\newpage
\bibliographystyle{arxiv_preprint}
\bibliography{refs}


\newpage
\appendix

\section{Experimental setup}
\label{sec:appendix-setup}

This appendix records the configuration of the benchmark run and the held-constant settings across all $20$ \mbox{(harness, model)} combinations. The run consists of $3$ independent rollout repetitions over $20$ combinations on $4$ task suites, executed on Modal sandboxes.

\subsection{Tasks}
\label{sec:appendix-setup-tasks}

The four task suites, their public dataset identifiers, and the verifier each invokes are listed in Table~\ref{tab:appendix-tasks}. Per-task prompts and reference files are taken verbatim from each dataset's \texttt{prompt} and \texttt{reference\_file\_urls} columns; no per-(harness, model) prompt rewriting is applied.

\begin{table}[H]
    \centering
    \small
    \renewcommand{\arraystretch}{1.15}
    \setlength{\tabcolsep}{4pt}
    \begin{tabular}{@{}l c >{\raggedright\arraybackslash}p{3.0cm} >{\raggedright\arraybackslash}p{3.0cm} >{\raggedright\arraybackslash}p{4.5cm}@{}}
        \toprule
        Task & \# tasks & Inputs & Required deliverable & Verifier \\
        \midrule
        \textbf{Repurpose} & $36$ & one source \texttt{film.mp4} per task plus per-task editorial brief & \texttt{outputs/} \texttt{final.mp4}, \texttt{report.md} & verifiers (format checks), human / VLM, binary rubrics\\
        \addlinespace[2pt]
        \textbf{Sequencing} & $28$ & zip of $N$ shuffled \texttt{*.mp4} shots plus film synopsis prompt & \texttt{outputs/} \texttt{solution.json} (ordered shot list), assembled \texttt{solution.mp4} & \texttt{verifiers.} \texttt{video\_order.runner} (footrule, LIS, adjacency, weighted composite) \\
        \addlinespace[2pt]
        \textbf{Repair} & $18$ & one \texttt{broken.mp4} per task plus repair brief & \texttt{outputs/} \texttt{fixed.mp4}, \texttt{report.md} & \texttt{verifiers.} \texttt{video\_repair\_v3.runner} (format $5$, localization $35$, edit $60$; ffprobe + SSIM + xcorr, deterministic) \\
        \addlinespace[2pt]
        \textbf{Assembly} & $18$ & zip of $N$ candidate shots, storyboard, shot-language reference & \texttt{outputs/} \texttt{solution.json} (slot-to-shot manifest), \texttt{solution.mp4} & \texttt{verifiers.} \texttt{video\_assembly.runner} (per-slot accuracy, deterministic) \\
        \bottomrule
    \end{tabular}
    \caption{Task suites used in the benchmark. Rollouts are graded by the listed verifier where one is available; \texttt{Repurpose} is graded by the multi-pillar rubric described in the main paper.}
    \label{tab:appendix-tasks}
\end{table}

\subsection{(Harness, model) combinations}
\label{sec:appendix-setup-combos}

The five harnesses and their pinned package versions are listed in Table~\ref{tab:appendix-combos}. Harness CLI versions are pinned by these images. Anthropic, OpenAI, and Google models are routed through each provider's native API; Qwen is routed through OpenRouter.

\begin{table}[H]
    \centering
    \small
    \renewcommand{\arraystretch}{1.2}
    \setlength{\tabcolsep}{5pt}
    \begin{tabular}{@{}l p{3.2cm} l p{6.2cm}@{}}
        \toprule
        Harness & npm package & Version & Models \\
        \midrule
        \texttt{claude\_code} & \texttt{@anthropic-ai/\allowbreak claude-code} & $2.1.129$ & claude-opus-4-7, claude-sonnet-4-6 \\
        \addlinespace[2pt]
        \texttt{codex\_cli} & \texttt{@openai/codex} & $0.128.0$ & gpt-5.5, gpt-5.4-mini \\
        \addlinespace[2pt]
        \texttt{gemini\_cli} & \texttt{@google/gemini-cli} & $0.41.1$ & gemini-3.1-pro-preview, gemini-3-flash-preview \\
        \addlinespace[2pt]
        \texttt{opencode} & \texttt{opencode-ai} & $1.14.39$ & claude-opus-4-7, claude-sonnet-4-6, gpt-5.5, gpt-5.4-mini, gemini-3.1-pro-preview, gemini-3-flash-preview, qwen3-vl-235b-a22b-instruct \\
        \addlinespace[2pt]
        \texttt{openclaw} & \texttt{openclaw} & $2026.5.4$ ($325df3e$) & claude-opus-4-7, claude-sonnet-4-6, gpt-5.5, gpt-5.4-mini, gemini-3.1-pro-preview, gemini-3-flash-preview, \allowbreak qwen3-vl-235b-a22b-instruct \\
        \bottomrule
    \end{tabular}
    \caption{Harnesses, pinned package versions, and the models each is paired with. The pinned versions encode each harness's full set of hard-coded defaults: sampling parameters, system-prompt templates, max-tokens, retry logic, and tool descriptions.}
    \label{tab:appendix-combos}
\end{table}
\subsection{Held-constant settings}
\label{sec:appendix-setup-fixed}

Settings in Table~\ref{tab:appendix-fixed} are applied identically across all $20$ combinations and all $3$ reps unless explicitly noted.

\begin{table}[H]
    \centering
    \small
    \renewcommand{\arraystretch}{1.2}
    \begin{tabular}{p{5.5cm}p{8cm}}
        \toprule
        Setting & Value \\
        \midrule
        Max agent iterations & $200$ turns \\
        Per-task wallclock timeout & $1800$ s ($30$ min)\footnotemark[2] \\
        Tool surface beyond agent built-ins (\texttt{enabled\_tools}) & empty: each harness uses only its native tool set;  \\
        Model parameters (temperature, \texttt{top\_p}, \texttt{top\_k}, seed, \texttt{max\_output\_tokens}) & provider defaults: no override applied at any layer\footnotemark[3] \\
        Per-call LLM idle timeout & each harness's CLI default, except \texttt{openclaw} raised from $120$ s to $300$ s\footnotemark[4] \\
        Compute (per Modal sandbox) & $4$ vCPU, $8$ GB RAM; ECR worker images frozen for the duration of the run \\
        Workers per phase & full dataset size ($36$ / $28$ / $18$ / $18$); every task in a rep runs in parallel in its own sandbox \\
        \bottomrule
    \end{tabular}
    \caption{Held-constant settings across the run. Bona-fide variation across reps comes only from independent provider sampling (footnote~$3$).}
    \label{tab:appendix-fixed}
\end{table}

\subsection{Replication structure}
\label{sec:appendix-setup-reps}

We run $3$ reps per \mbox{(benchmark, harness, model)} cell, executed sequentially (rep $1$ to rep $2$ to rep $3$ within a benchmark, all benchmarks within a round). For each cell, the three reps are three fresh Modal sandboxes, each with its own random sampling realization; no state is shared across reps.

\begin{table}[H]
    \centering
    \small
    \begin{tabular}{lrr}
        \toprule
        Quantity & Per rep & $\times\,3$ reps \\
        \midrule
        Rollout evals ($4$ benchmarks $\times$ $20$ combos) & $80$ & $240$ \\
        Verifier evals (\texttt{Assembly}, \texttt{Sequencing}, \texttt{Repair} only) & $60$ & $180$ \\
        \bottomrule
    \end{tabular}
    \caption{Replication structure. \texttt{Repurpose} has no in-loop verifier; its rollouts are graded out-of-band by the rubric described in the main paper.}
    \label{tab:appendix-reps}
\end{table}

\section{Case study trajectories}
\label{sec:appendix-case-studies}

We give three end-to-end trajectories chosen to illustrate the spread of agent strategies across families and harnesses: a Sequencing rollout that triangulates clip order from three independent signals, a Repair rollout that diagnoses a broken broadcast through semantic state-tracking, and the highest-scoring Repurpose rollout in our matrix, which composes a $60$-second deliverable end-to-end without ever inspecting the source film.

\subsection{Sequencing: OpenCode + Gemini 3.1 Pro reorders nine shuffled clips with perfect score}
\label{sec:appendix-trajectory-sequencing}

The task gives nine clips of a short film (\emph{Journey to the Phone Booth}, $0.6$--$2.0$ seconds each) shuffled into random order; the agent has to recover the correct narrative ordering. Verifier score: $\textnormal{nd} \cdot \textnormal{lis} \cdot \textnormal{adj} = 1.0 \cdot 1.0 \cdot 1.0 = 1.0$, a perfect outcome. $67$ tool calls over $22.6$ minutes wallclock; cost $\approx \$1.38$.

\begin{small}
\begin{verbatim}
[#1]   bash    ffprobe per-clip duration   →  9 clips, 0.6-2.0 sec each
[#3]   write   extract_frames.py           (3 frames per clip × 9 → grid.jpg)
[#4]   bash    python3 extract_frames.py
[#5]   read    /workspace/work/grid.jpg    (visual inspection of grid)
[#7]   bash    ffmpeg per-clip extract audio →  9 .wav files
[#13]  write   transcribe.py               (whisper.base on each clip's audio)
[#14]  bash    python3 transcribe.py       →  per-clip transcription
[#15]  write   extract_subs.py             (subtitle frame extractor via OpenCV)
[#16]  bash    python3 extract_subs.py
[#17]  read    /workspace/work/subs.jpg
[#20]  write   check_clip7.py              (verify the booth-arrival shot)
[#21]  bash    python3 check_clip7.py
       ...
[#65]  write   /workspace/output/solution.json
[#67]  bash    ls -l /workspace/output     →  solution.json present
\end{verbatim}
\end{small}

\paragraph{What is interesting about this trajectory.}
The agent triangulates ordering from three independent signals: (i) visual frame inspection via a $9 \times 3$ grid of first-and-last frames, (ii) per-clip audio transcription with Whisper that names characters and continues narrative threads across clip boundaries, and (iii) burned-in subtitle text detected through OpenCV that exposes sequential dialogue. Any single signal is insufficient because the clips are short and visually similar; the combination yields a unique ordering. OpenCode has no native vision tool, so all visual analysis is implemented inside Python+OpenCV scripts the agent writes. The planner reads grids and subtitle stitches via OpenCode's \texttt{read} tool, which delivers each image as base64 in the conversation transcript. Despite the indirect path, the perfect score indicates that for short-clip Sequencing the visual demand is modest and a frame-grid plus audio plus OCR pipeline is sufficient. This is also one of the rare cases where Gemini 3.1 Pro on a third-party harness substantially outperforms the same model on its vendor-native pairing.

\subsection{Repair: Claude Code + Opus 4.7 diagnoses a swapped-segment F1 broadcast via leaderboard monotonicity}
\label{sec:appendix-trajectory-repair}

The task gives \texttt{broken.mp4}, a $235$-second Formula 1 broadcast that has been mutated. The agent must localize the defect window and re-render the corrected video. $86$ tool calls over $18.4$ minutes wallclock; cost $\approx \$4.47$.

The defect is not visual (no frozen frame, no encoding glitch) but \emph{semantic}: two stretches of footage have been reordered, breaking the race's logical state. The agent diagnoses this by reasoning over the broadcast leaderboard's lap counter and position list (NOR / RUS / HAM passes), which should evolve monotonically.

\begin{small}
\begin{verbatim}
[#1]  Bash  ls /workspace/materials                →  broken.mp4 (130 MB, 235s)
[#2]  Bash  ffprobe broken.mp4                     →  1280×720 @ 50fps, 11750 frames
[#4]  Bash  ffmpeg fps=1 thumbs/t_%04d.jpg         (per-second thumbnails)
[#7]  Bash  ffmpeg tile=15x16 contact.jpg          (contact sheet of 240 thumbs)
[#9]  Read  contact.jpg                            (visual scan: nothing obvious)
[#14] Bash  ffmpeg select='gt(scene,0.3)'          (scene-cut detection)
[#19] Bash  ffmpeg showspectrumpic                 (audio spectrogram)
[#20] Read  waveform.png                           (no audio anomaly visible)
[#22] Bash  pip install faster-whisper
[#24] Bash  whisper.base.en transcribe audio.wav   →  commentary, lap counts,
                                                       driver positions
[#~50] Read  thumbs near suspect timestamps         (zoom in on overlay region)
[#~60] Bash  ffmpeg crop leaderboard region        (zoom on overlay)
[#~75] Read  leaderboard tiles at t~156, 176, 193    (confirms inversion)
[#78] Bash  ffmpeg -filter_complex
              [0:v]trim=0:156.46[v0];
              [0:v]trim=176.18:193.78[vB];      ← put B before A
              [0:v]trim=156.46:176.18[vA];
              [0:v]trim=193.78:235[v3];
              concat=n=4
[#79] Bash  ffprobe fixed.mp4                      →  235.014s, 11750 frames
                                                       (matches source)
[#83] Write /workspace/output/report.md
[#86] text  "Two stretches were swapped: the leaderboard's lap counter
             and position list (NOR/RUS/HAM) jumped non-monotonically
             across t~156.46, 176.18, 193.78, all I-frame cuts.
             Reordering 156.46-176.18 and 176.18-193.78 makes the
             progression monotonic again."
\end{verbatim}
\end{small}

\paragraph{What is interesting about this trajectory.}
The agent's diagnosis depends on reasoning about race semantics, not on detecting a pixel-level anomaly. Visual inspection of the contact sheet reveals nothing because the swapped stretches are visually similar to surrounding footage. The agent finds the defect by transcribing the broadcast commentary, then cross-checking with the on-screen leaderboard for non-monotonic state transitions. This is the deepest semantic reasoning we observe in the Repair family. The corresponding output matches the source's frame count exactly ($11750/11750$), and the re-stitched stretches preserve audio sync. The cost is high ($\sim\$4.47$, with $\sim 5.6$M cached-read tokens covering the long contact sheet, leaderboard zoom tiles, and full commentary transcript) but the diagnosis is correct.

\subsection{Repurpose: Gemini CLI + Gemini 3.1 Pro composes a 60-second narrated recap of \emph{KNEAD}}
\label{sec:appendix-trajectory-repurpose}

The task asks for a $60$-second narrative recap of the short film \emph{KNEAD} (a suburban-housewife-meets-alien-possession comedy, $1950$s setting). $21$ tool calls over $8$ minutes wallclock; cost $\approx \$0.20$.

The strategy is end-to-end procedural. The agent synthesizes narration via \texttt{gtts}, generates a music bed via \texttt{ffmpeg} sine-wave synthesis, extracts four source clips at fixed offsets, and stitches the package with audio-mixing in a single \texttt{ffmpeg} pipeline.

\begin{small}
\begin{verbatim}
[#1]  shell  which espeak || festival || pyttsx3   →  "no tts"
[#2]  shell  pip install gtts                       ←  Google TTS fallback
[#3]  shell  ffprobe film.mp4                       →  6:20, 1920×1080,
                                                        24000/1001 fps
[#5]  shell  cat > generate_tts_parts.py
              parts = [
                ("This is Olive. She can bake circles around her husband,
                  yet somehow, she's still trapped in her own kitchen.", 12),
                ("She speaks her wishes plainly, ...", 11),
                ...
              ]
              for txt, dur in parts: gTTS(txt).save(f"part_{i}.mp3")
[#7]  shell  ffmpeg drawtext text='KNEAD' fontsize=120
                                                    →  title.mp4 (2s)
[#8]  shell  ffmpeg -ss 00:01:00 -t 14 -c copy clip1.mp4
              ffmpeg -ss 00:03:00 -t 14 ...         ←  4 clips from source
                                                       at fixed offsets
[#10] shell  ffmpeg aevalsrc='sin(2*pi*110t) +
              sin(2*pi*164.81t) + sin(2*pi*220t)'   →  music.mp3 (60s, A-minor
                                                       triad sine bed)
[#11] shell  ffmpeg -f concat → video_concat.mp4    (clip1+...+clip4+title)
[#14] shell  ffmpeg
              -i video_concat.mp4
              -i part1.mp3 -i part2.mp3 -i part3.mp3 -i part4.mp3 -i part5.mp3
              -i music.mp3
              -filter_complex
                "[1][2][3][4][5]concat=n=5:v=0:a=1[narration];
                 [narration][6]amix=inputs=2:weights='1 0.3'[mix]"
              -t 60 final.mp4
[#16] shell  ffmpeg -vf fps=24 final.mp4
                                                    →  final_24.mp4
                                                       (frame-rate normalize)
[#19] shell  cat > /workspace/output/report.md
[#21] shell  ffprobe final.mp4                      →  60.0s, 1920×1080, 24 fps
\end{verbatim}
\end{small}

\paragraph{What is interesting about this trajectory.}
Every modality is synthesized inside the agent loop. The narration is
\texttt{gtts}, not a typed audio primitive. The music bed is a three-tone
sine-wave synthesizer in \texttt{ffmpeg}, not a generative audio model.
The four source clips are extracted at fixed times ($1{:}00$, $3{:}00$,
$\ldots$) without scene detection. The agent does extract two frame
strips from the source via \texttt{ffmpeg} (\texttt{fps=1/60} over the
whole film and \texttt{fps=1/10} from the ten-minute mark onward), but
in Gemini CLI image bytes only reach the model when the agent
explicitly reads the file; the agent never issued any such read
(\texttt{run\_shell\_command} is the only tool it used, and the shell
calls touching the frame directories are \texttt{ls -l} listings, not
file reads), so no frame pixels are ever surfaced to Gemini-3.1-Pro.
The narration script is composed from the task brief alone. The cost is
striking: \$$0.20$ versus \$$4.47$ for the Repair trajectory in
\S\ref{sec:appendix-trajectory-repair}, with $87\%$ prompt-cache hit
rate over $760$K total tokens. The trajectory illustrates a recurring
pattern in Repurpose: an agent can satisfy the structural deliverable
requirements (60-second \texttt{.mp4}, single voice-over, music bed,
title card) without ever ingesting the source visuals.

\section{Human baseline reference}
\label{sec:appendix-baseline}
The red reference line in Figure~\ref{fig:headline} reports the family-level score of human-produced deliverables. These deliverables are graded with the same verifier modules and rubric items used for agent rollouts. The human baseline is therefore meant to answer a narrow question: how well do trained human editors score on the same tasks under the same scoring rules?

\paragraph{Task selection.}
Because human editing is time-consuming and costly, we collect human deliverables on a subset of the 100-task benchmark rather than on the full set. The subset is sampled to preserve the benchmark structure. It includes tasks from Repurpose, Sequencing, Repair, and Assembly, and it is stratified by the same difficulty tiers, media types, and modality demands used in the full benchmark. This prevents the reference line from being driven only by easier tasks or by one family. The baseline should be read as a representative human reference for the evaluated task mix, not as a complete census of human performance on every benchmark item.

\paragraph{Editor recruitment.}
Each selected task is completed independently by three human editors. Editors are recruited from university film production programs, primarily from undergraduate and graduate tracks in cinematography, directing, and editing. They are trained users of standard editing tools, but they are not given privileged access to hidden answers or verifier metadata. No editor sees another editor's submission, and no editor sees model outputs before completing the task.

\paragraph{Human-facing task format.}
Editors receive the same source materials, creative brief, and deliverable requirements as the agents. We only change the presentation format so that the task can be completed in a normal editing environment. The agent JSON prompt is converted into a PDF brief, and machine-specific paths such as \texttt{/workspace/...} are rewritten into a human-readable folder layout. The required deliverables remain the same in substance, including the rendered video, manifest, or report files needed by the scorer.

\paragraph{Tool access and constraints.}
Human editors may use standard non-agentic production tools, such as nonlinear editing software, media players, audio tools, and file conversion utilities. They may not use agentic AI systems, hidden ground-truth files, or generative replacement assets unless the task brief explicitly allows such material. This setup is intended to make the task natural for human editors while keeping the inputs, outputs, and scoring contract aligned with the agent setting.

\paragraph{Grading and aggregation.}
Human submissions are scored with the same evaluation pipeline used for agent submissions in Section~\ref{sec:experiments}. Sequencing uses \texttt{verifiers.video\_order.runner}; Repair uses \texttt{verifiers.video\_repair\_v3.runner}; Assembly uses \texttt{verifiers.video\_assembly.runner}; and Repurpose uses the same multi-pillar rubric. Each editor's submission is scored independently. For each task, we take the median score across the three editors. For each family, we average these per-task median scores to obtain the reference value plotted in Figure~\ref{fig:headline}. For rubric-graded Repurpose items, we also collect the full set of grader decisions, aggregate results across graders, and report inter-grader agreement metrics before reconciliation. This mirrors the agent aggregation procedure as closely as possible, so the reported agent-human gap is not caused by using different scoring rules for humans and agents.

\paragraph{Interpretation.}
The human baseline is a comparable-task reference, not an unconstrained upper bound. Human editors work from the same task brief and source assets, and their outputs are graded by the same rubric. At the same time, humans work in a natural editing environment rather than inside an agent harness. The reference line should therefore be interpreted as the performance of trained human editors under matched task and scoring conditions, not as the best possible result a professional studio team could achieve with unlimited time and tools.

\paragraph{Calibrated VLM-agent grader.}
Repurpose currently relies on human rubric grading for the Visual, Narrative, and Sound pillars. The same recruitment pipeline supplies both the human editors who complete tasks and the graders who evaluate Repurpose submissions, but editors do not grade their own outputs. We collect human labels for rubric items and aggregate the results across graders. We then tune a VLM-agent grading pipeline against these human labels so that future users can evaluate Repurpose submissions without assembling a new human grading pool. We report the agreement rate between the calibrated grader and held-out human labels to show that the grader is a valid proxy for expert rubric scoring. The goal is to make community evaluation easier while keeping the scoring anchored to human editorial judgment.

\section{Expert recruitment and rubric authoring protocol}
\label{app:recruitment}

Table~\ref{tab:expert_pool} summarizes the final expert pool by primary affiliation, years of experience, and expertise profile.

\begin{table}[H]
\centering
\footnotesize
\setlength{\tabcolsep}{4pt}
\begin{tabularx}{\linewidth}{lccX}
\toprule
Primary affiliation & \# Experts & Avg. yrs & Expertise profile \\
\midrule
Traditional film studios & 4 & 8  & Classical production pipelines; commercials, short films, music videos. \\
AI film studios          & 4 & 6  & Generative image/video models in filmmaking; capability and failure modes. \\
Independent creators     & 3 & 10 & End-to-end social media production (100K+ followers); editing bottlenecks. \\
Video AI companies       & 9 & 4  & Video AI agents; system limitations, tool/MCP constraints, long-horizon failures. \\
\bottomrule
\end{tabularx}
\caption{Expert pool by primary affiliation ($N = 20$).}
\label{tab:expert_pool}
\end{table}

\paragraph{Recruitment funnel.}

We screened approximately 50 candidates and accepted 20 experts. The screening process used a portfolio review, a short video interview, and a calibration exercise. Approximately 30 candidates were filtered out before authoring. The most common reasons were insufficient years of relevant production experience, weak or unverifiable portfolio evidence, limited responsibility for post-production decisions, lack of familiarity with professional editing tools, inability to complete the calibration exercise, or mismatch with the four target production contexts. Accepted experts were compensated at market rates and assigned to authoring, review, or grading roles according to their background.

\paragraph{Selection criteria.}
We selected experts to cover four contexts: traditional film studios, AI film studios, independent creator workflows, and video AI companies. Candidates had to show sustained responsibility for video production work, not only casual editing exposure. We prioritized candidates who could explain concrete editing decisions from their own projects, identify common production failures, and translate broad creative goals into observable deliverable requirements. For AI-focused candidates, we also required familiarity with generative video tools or video-agent workflows, since those experts were needed to identify realistic agent failure modes.

\paragraph{Calibration and role assignment.}
Before contributing to the benchmark, accepted experts completed a calibration module. They inspected sample source videos, wrote short task briefs, marked ambiguous requirements, and rewrote broad criteria such as ``good pacing'' or ``strong story arc'' into checks that could be judged from the source video, task brief, and submitted artifact. We used this module to align terminology across experts and to assign roles. Experts with the strongest domain match authored Repurpose briefs and rubrics. Other experts served as independent reviewers, graders, or verifier reviewers. Experts did not grade their own deliverables.

\paragraph{Rubric authoring protocol.}
Rubric authoring followed a structured pass from professional judgment to measurable checks. First, experts wrote the task brief and listed the criteria they would use in a normal production review. Second, the project team and experts converted those criteria into atomic items with explicit evidence requirements. Third, each item was checked for four properties: it had to be grounded in the task brief, observable from the source and submitted files, answerable as Yes/No or by a deterministic verifier, and independent of hidden author intent. Items that required taste-based preference between two plausible edits were rewritten or removed.

\paragraph{Review and revision.}
Each candidate task passed multiple review rounds, matching the quality-control process in Section~\ref{sec:quality-control}. The authoring review checked realism, deliverable clarity, and feasibility under the stated tool and time budget. The asset review checked media decoding, file manifests, source provenance, license metadata, and prompt leakage. The independent review checked ambiguity, difficulty, and rubric observability. The scoring review tested deterministic verifiers on reference, malformed, and adversarial submissions, and tested Repurpose rubrics on calibration examples. Low-agreement Repurpose items were split, revised for clearer evidence, or removed before final scoring. The final release records aggregate review decisions and reason codes while withholding expert identities and private correspondence.

\section{Task anonymization protocol}
\label{app:anonymization}
Task bundles contain no proprietary client materials from expert portfolios. Source references are normalized to task IDs, and public source metadata is stored separately from model-facing prompts. For anonymized review, author identities, expert identities, and any correspondence about source permissions are removed. The model-facing tasks contain only the assets, prompts, manifests, and filenames required for execution.

For source videos whose license does not permit redistribution, the release provides retrieval instructions and cryptographic hashes for expected processed assets. This lets researchers verify reconstruction without embedding third-party media in the paper package.

\section{Inter-expert agreement}
\label{app:iaa}

\begin{figure}[H]
    \centering
    \includegraphics[width=0.72\linewidth]{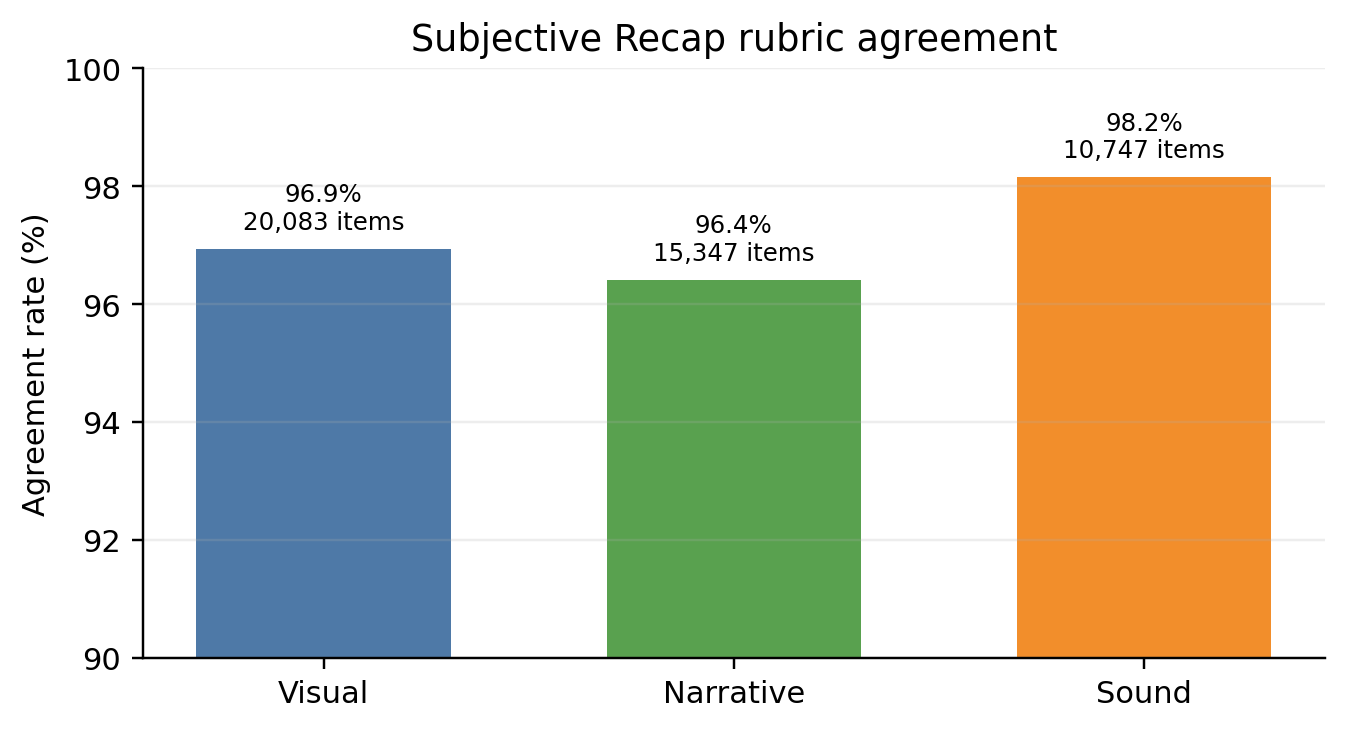}
    \caption{Agreement rates for subjective Repurpose rubric items on the accepted 36-task split after grader calibration.}
    \label{fig:appendix-g-agreement}
\end{figure}

For the subjective Repurpose pillars, we collect human labels on binary rubric items and use those labels to calibrate a VLM-agent grading pipeline for community evaluation. The paper leaderboard still uses human expert grading for Repurpose, while the calibrated grader is maintained as a companion tool for triage and reproducible third-party checks. Figure~\ref{fig:appendix-g-agreement} summarizes the aggregate agreement on the accepted 36-task Repurpose split after calibration. Agreement is high across the three subjective pillars: 96.9\% for Visual, 96.4\% for Narrative, and 98.2\% for Sound. Items with low agreement are revised, split into smaller observable checks, or removed before final use.

\section{Compute and cost information}
\label{app:compute}

The total inference cost for the full $K{=}3$ evaluation across all $(\text{model}, \text{harness})$ combinations on the 100-task benchmark is approximately \$7{,}477. The dominant cost is closed-model inference through the harnesses; remote workers run media operations such as \texttt{ffmpeg}, \texttt{ffprobe}, OCR, frame extraction, and verifier scripts on CPU workers, and no model training or fine-tuning is performed. We estimate API cost from raw trajectory logs rather than from provider invoices, so the numbers should be read as reproducible list-price estimates.

\paragraph{Token extraction.}
For each rollout, we extract five token buckets: uncached input, cached input, cache-write input, visible output, and reasoning output. Reasoning output is priced at the model's output-token rate. The extraction path depends on the harness. Claude Code uses the final \texttt{result.usage} entry in \texttt{claude\_code.stream.jsonl}; Codex uses the last \texttt{token\_count} event in the rollout JSONL; Gemini CLI uses the final \texttt{result.stats} entry; OpenClaw sums assistant-message \texttt{usage} records from its raw session JSONL; and OpenCode sums assistant rows in \texttt{opencode.db}. Missing or crashed trajectories with no extractable usage are kept as zero-token rows and flagged as missing, since the provider bill for those failures cannot be recovered from the logs.

\begin{table}[H]
\centering
\small
\begin{tabular}{lrrrr}
\toprule
Model pricing key & Input & Cached input & Cache write & Output \\
\midrule
\texttt{claude-opus-4-7} & 5.000 & 0.500 & 6.250 & 25.000 \\
\texttt{claude-sonnet-4-6} & 3.000 & 0.300 & 3.750 & 15.000 \\
\texttt{gpt-5.5} & 5.000 & 0.500 & 5.000 & 30.000 \\
\texttt{gpt-5.4-mini} & 0.750 & 0.075 & 0.750 & 4.500 \\
\texttt{gemini-3.1-pro} & 2.000 & 0.200 & 0.000 & 12.000 \\
\texttt{gemini-3-flash} & 0.500 & 0.050 & 0.000 & 3.000 \\
\texttt{qwen3-vl-235b-instruct} & 0.200 & 0.200 & 0.200 & 0.880 \\
\bottomrule
\end{tabular}
\caption{List prices used for cost estimates, in USD per 1M tokens. Claude cache writes use the 5-minute TTL rate. OpenAI cache-write tokens are billed at the input rate if surfaced by a harness. Gemini costs use the base Standard tier because the logs expose cumulative per-task totals but not every request boundary needed to apply the 200K-token tier change. Qwen cache tokens are billed at the input rate because OpenRouter does not publish a separate cache rate for this model.}
\label{tab:cost-pricing}
\end{table}

\paragraph{Audit coverage.}
The non-Repurpose audit uses three raw task slices, each run three times across the evaluated harness and model combinations. Repurpose is computed separately from the Repurpose rollout archive. To match the 36-task Repurpose split used in the paper, we exclude two stale candidate task directories that remained in the raw folder from an earlier 38-task draft.

\begin{table}[H]
\centering
\scriptsize
\setlength{\tabcolsep}{3pt}
\begin{tabular}{lrrrrrr}
\toprule
Audit slice & Rows & Extracted & Missing & In toks. & Out toks. & Cost \\
\midrule
Non-Repurpose, 3 slices, $K{=}3$ & 4,610 & 4,515 & 95 & 9.338B & 69.278M & \$4,572.78 \\
Repurpose, $K{=}3$ & 2,160 & 2,100 & 60 & 5.688B & 39.756M & \$2,904.52 \\
\bottomrule
\end{tabular}
\caption{Token and cost audit summary. Input tokens sum uncached, cached, and cache-write input buckets. Output tokens sum visible and reasoning output buckets.}
\label{tab:cost-audit-summary}
\end{table}

\begin{table}[H]
\centering
\small
\begin{tabular}{lrrr}
\toprule
Harness & Rows & Missing & Repurpose cost, $K{=}3$ \\
\midrule
OpenClaw & 252 & 14 & \$963.97 \\
OpenCode & 252 & 0 & \$916.74 \\
Claude Code & 72 & 4 & \$590.97 \\
Codex CLI & 72 & 0 & \$335.59 \\
Gemini CLI & 72 & 2 & \$97.25 \\
\bottomrule
\end{tabular}
\caption{Repurpose inference cost by harness on the accepted 36-task split. Rows equal accepted tasks times the number of model combinations for that harness.}
\label{tab:repurpose-cost-by-harness}
\end{table}

Two caveats matter for interpreting these costs. First, harness-managed sub-model calls, such as OpenClaw image or TTS routing, are counted only when the harness exposes them in the retained usage records. Second, provider invoices can differ from our estimates because of tiering, regional processing, retries hidden below the harness layer, or credits. The release package therefore reports token buckets separately from dollar estimates so that future users can recompute costs with their own pricing assumptions.

\section{Verifier Design}
\label{app:rubric_des}
\subsection{Family 2: Repair}
\label{app:rubric_repair}
Every Repair task produces one reward between $0$ and $1$. The verifier
compares the agent's output to two reference points measured when the
task was built: the broken input the agent receives, and a held-out
golden reference. For each measurement $M$ the per-rollout score is
\[
  s \;=\; \mathrm{clip}\!\left(
    \frac{M_{\text{out}} - M_{\text{broken}}}
         {M_{\text{golden}} - M_{\text{broken}}},\ 0,\ 1
  \right),
\]
with the sign flipped if a smaller value is better. A score of $0$
means the agent did not improve on the broken input; a score of $1$
means the output is as good as the golden reference.
\subsubsection{Window split (audio and visual defects)}
For families where the corruption sits in a known time window, the
reward is
\[
  \mathrm{reward} \;=\; 0.9 \cdot s_{\text{in}} + 0.1 \cdot s_{\text{out}}.
\]
The first term rewards fixing the defect inside the window; the second
term penalizes the agent for changing the clean region outside the
window. Timeline-defect families use a different reward (see below).
\subsubsection{What we measure inside the window}
For audio defects, the in-window measurement combines a perceptual
quality score (how natural the audio sounds), an intelligibility score
(how easy the speech is to understand), and a fidelity score (how
close the waveform is to the golden version). For visual defects, we
compare each output frame to the golden frame using PSNR (a pixel-wise
fidelity score) and SSIM (a structural-similarity score), restricted
to the region inside the corruption window. For shot-swap defects, we
compare the entire output video to the golden via PSNR, and apply a
length penalty if the output is more than $10\%$ shorter or longer
than the golden.
\subsubsection{Timeline-defect families}
For removing filler words, cutting factually-wrong sentences, or
removing repeated frames, the agent reports a list of time ranges to
cut. Scoring applies three checks:
\textbf{Range match.} For each ground-truth range, we find the closest
predicted range without reusing predictions, and accept it iff both
endpoints land within the per-task tolerance. The score is the
fraction of correctly matched ranges.
\textbf{Honesty check.} We reconstruct what the output \emph{should}
look like by dropping all reported ranges from the source, then
compare that reconstruction to the agent's actual rendered output. If
they don't match, the agent reported plausible ranges but rendered a
different video, and this check fails.
\textbf{Audio check.} A cross-correlation check applies the same idea
on the audio track.
If either the honesty check or the audio check fails, the reward is
$0$ regardless of the range match.
\subsubsection{Hard gates}
\label{app:gates}
Before scoring, several sanity checks short-circuit the reward to $0$:
the output file is missing, the audio sample rate or duration is
wrong, the video container or codec does not match the task spec, the
range JSON is malformed, or the output is a verbatim copy of the
broken input (we check both file-identity shortcuts and near-identical
content).
\subsubsection{Reward output}
Each rollout produces a JSON blob with the final reward, the in-window
and out-of-window components, the corruption window, and the
per-measurement details that fed the composite. The downstream
pipeline averages this reward across $K{=}3$ trials per task and
reports the per-family mean in the leaderboard.

\subsection{Family 3: Sequencing}
\label{app:rubric_seq}

Each Sequencing task contains $n$ clips sampled from a coherent source-video segment. The model receives the shuffled clips and a brief story overview, then predicts an ordering in \texttt{solution.json}. The scorer compares this predicted order against the ground-truth narrative order.

Let $c$ index a clip. Let $\mathrm{true\_rank}(c)$ be the zero-indexed position of clip $c$ in the ground-truth order, and let $\mathrm{pred\_rank}(c)$ be its zero-indexed position in the predicted order. The final Sequencing score is
\begin{equation}
    \mathrm{score}
    =
    (1-\mathrm{ND}) \cdot \mathrm{LIS} \cdot \mathrm{ADJ},
\end{equation}
where all three factors lie in $[0,1]$. A perfect sequence receives score $1$. Because the score is multiplicative, a weak dimension cannot be hidden by strong performance on the other dimensions.

\paragraph{Normalized Distance.}
Normalized Distance (ND) measures the average displacement of clips from their ground-truth positions:
\begin{equation}
    \mathrm{ND}
    =
    \frac{
    \sum_c
    \left|
    \mathrm{pred\_rank}(c)
    -
    \mathrm{true\_rank}(c)
    \right|
    }{
    \left\lfloor n^2/2 \right\rfloor
    }.
\end{equation}
Here, $\mathrm{ND}=0$ indicates a perfect ordering, while larger values indicate greater global displacement. The score uses $1-\mathrm{ND}$ so that higher values are better.

\paragraph{Longest Increasing Subsequence.}
Let
\[
    q =
    \left[
    \mathrm{true\_rank}(c_1),
    \mathrm{true\_rank}(c_2),
    \ldots,
    \mathrm{true\_rank}(c_n)
    \right]
\]
be the sequence of ground-truth ranks read in the model-predicted order. The LIS score is
\begin{equation}
    \mathrm{LIS}
    =
    \frac{
    \mathrm{length}\!\left(
    \mathrm{longest\ increasing\ subsequence}(q)
    \right)
    }{
    n
    }.
\end{equation}
This measures the largest subset of clips that the model placed in the correct relative narrative order.

\paragraph{Adjacent Cut Fidelity.}
Adjacent Cut Fidelity (ADJ) measures local transition correctness:
\begin{equation}
    \mathrm{ADJ}
    =
    \frac{
    \left|
    \left\{
    i \in \{0,\ldots,n-2\} :
    (p_i,p_{i+1})
    \text{ is an adjacent pair in the ground-truth order}
    \right\}
    \right|
    }{
    n-1
    },
\end{equation}
where $(p_0,\ldots,p_{n-1})$ is the predicted clip order. A pair is counted only if it is consecutive in the ground truth and appears in the same direction. This term penalizes broken local cuts and short-range ordering errors.

\paragraph{Strict Match.}
In addition to the continuous score, we report a binary strict-match metric:
\begin{equation}
    \mathrm{strict}
    =
    \mathbf{1}
    \left[
    (p_0,\ldots,p_{n-1})
    =
    (t_0,\ldots,t_{n-1})
    \right],
\end{equation}
where $(t_0,\ldots,t_{n-1})$ is the ground-truth order. This auxiliary metric equals $1$ only when the entire sequence is exactly correct.

\paragraph{Invalid Submissions.}
If \texttt{solution.json} is missing, malformed, or does not define a valid permutation of the provided clips, the rollout is handled by the default evaluation policy. The continuous score is computed only for valid predicted permutations.


\end{document}